\documentclass[10pt,aps,prl,twocolumn,superscriptaddress]{revtex4-2}

\usepackage{newtxtext,newtxmath}

\usepackage[T1]{fontenc}
\usepackage{microtype}
\usepackage[pdftex, hidelinks]{hyperref}
\usepackage{xcolor}
\definecolor{linkcolor}{rgb}{0.0, 0.47, 0.75}
\hypersetup{%
  linkcolor=linkcolor,
  citecolor=linkcolor,
  urlcolor=linkcolor,
  colorlinks=true
}

\usepackage{csquotes}
\usepackage[american]{babel}
\usepackage{relsize}
\usepackage{tex/macros}
\usepackage{graphicx}
\graphicspath{{./img/}}
\usepackage{booktabs}
\usepackage{amsmath}
\usepackage{orcidlink}

  \renewcommand{\epsilon}{\varepsilon}
  \renewcommand{\theta}{\vartheta}
  \renewcommand{\rho}{\varrho}
  \renewcommand{\phi}{\varphi}
\usepackage[capitalise]{cleveref}

\begin{document}

\title{%
  First Results on the Search for Lepton Number Violating \\
  Neutrinoless Double Beta Decay with the \legend-200 Experiment
}

% authors-and-affiliations.tex
%May 8, 2025 version
\newcommand{\FootText}{Institutional Board membership suspended since April 26, 2022.}

% affiliations:

\newcommand{\MPP}{Max-Planck-Institut f\"{u}r Physik, Garching b. M\"{u}nchen, 85748, Germany}
\newcommand{\UNM}{Department of Physics and Astronomy, University of New Mexico, Albuquerque, NM 87131, USA}
\newcommand{\LAquila}{Dipartimento di Scienze Fisiche e Chimiche dell'Universit\`{a} degli Studi dell'Aquila, L'Aquila, 67100, Italy}
\newcommand{\GSSI}{Gran Sasso Science Institute, L'Aquila, 67100, Italy}
\newcommand{\LNGS}{Istituto Nazionale di Fisica Nucleare, Laboratori Nazionali del Gran Sasso, 67100 Assergi (AQ), Italy} 
\newcommand{\UTAustin}{Department of Physics, University of Texas at Austin, Austin, TX 78712, USA}
\newcommand{\LBNLNSD}{Institute for Nuclear and Particle Astrophysics and Nuclear Science Division, Lawrence Berkeley National Laboratory, Berkeley, CA 94720, USA}
\newcommand{\LBNLENG}{Engineering Division, Lawrence Berkeley National Laboratory, Berkeley, CA 94720, USA}
\newcommand{\UCBNE}{Department of Nuclear Engineering, University of California, Berkeley, CA, 94720, USA}
\newcommand{\IKZ}{Leibniz-Institut f\"{u}r Kristallz\"{u}chtung, Berlin, D-12489, Germany}
\newcommand{\IU}{Center for Exploration of Energy and Matter, and Department of Physics, Indiana University, Bloomington, IN 47405, USA}
\newcommand{\Bratislava}{Department of Nuclear Physics and Biophysics, Comenius University, Bratislava, SK-84248, Slovakia}
\newcommand{\SFU}{Department of Chemistry, Simon Fraser University, Burnaby, British Columbia, V5A 1S6, Canada}
\newcommand{\UNC}{Department of Physics and Astronomy, University of North Carolina, Chapel Hill, NC 27599, USA}
\newcommand{\Warwick}{Department of Physics, University of Warwick, Coventry, CV4 7AL, United Kingdom}
\newcommand{\TUNL}{Triangle Universities Nuclear Laboratory, Durham, NC 27708, USA}
\newcommand{\Duke}{Department of Physics, Duke University, Durham, NC 27708, USA}
\newcommand{\USC}{Department of Physics and Astronomy, University of South Carolina, Columbia, SC 29208, USA}
\newcommand{\Jag}{M. Smoluchowski Institute of Physics, Jagiellonian University, Cracow, 30-348, Poland}
\newcommand{\Dresden}{Technische Universit\"{a}t Dresden, Dresden, 01069, Germany}
\newcommand{\JINR}{Joint Institute for Nuclear Research, Dubna, 141980, Russia} 
\newcommand{\INRRAS}{Institute for Nuclear Research of the Russian Academy of Sciences, Moscow, 119991, Russia}
\newcommand{\Geel}{European Commission, Joint Research Centre, Directorate for Nuclear Safety \& Security, Geel, 2440, Belgium}
\newcommand{\MPIK}{Max-Planck-Institut f\"{u}r Kernphysik, Heidelberg, 69117, Germany}
\newcommand{\Queens}{Department of Physics, Engineering Physics \& Astronomy, Queen's University, Kingston, Ontario, K7L 3N6, Canada} 
\newcommand{\UTK}{Department of Physics and Astronomy, University of Tennessee, Knoxville, TN 37916, USA}
\newcommand{\Lancaster}{Department of Physics, Lancaster University, Lancaster, LA1 4YW, United Kingdom}
\newcommand{\Liverpool}{University of Liverpool, Liverpool, L69 3BX, United Kingdom}
\newcommand{\UCL}{Department of Physics and Astronomy, University College London, London, WC1E 6BT, United Kingdom}
\newcommand{\LANL}{Los Alamos National Laboratory, Los Alamos, NM 87545, USA}
\newcommand{\MILB}{Universit\`{a} degli Studi di Milano Bicocca, Milan, 20126, Italy}
\newcommand{\MILBINFN}{Istituto Nazionale di Fisica Nucleare, Sezione di Milano Bicocca, Milan, 20126, Italy}
\newcommand{\MILC}{Universit\`{a} degli Studi di Milano, Milan, 20133, Italy}
\newcommand{\MILCINFN}{Istituto Nazionale di Fisica Nucleare, Sezione di Milano, Milan, 20133, Italy}
\newcommand{\NRCKI}{National Research Centre ``Kurchatov Institute'', Moscow, 123098, Russia}
\newcommand{\MEPhI}{National Research Nuclear University MEPhI (Moscow Engineering Physics Institute), 115409 Moscow, Russia}
\newcommand{\TUMPhy}{Department of Physics, TUM School of Natural Sciences, Technical University of Munich, 85748 Garching b. M\"{u}nchen, Germany}
\newcommand{\ORNL}{Oak Ridge National Laboratory, Oak Ridge, TN 37830, USA}
\newcommand{\PadovaUniv}{Dipartimento di Fisica e Astronomia dell'Universit\`{a} degli Studi di Padova, Padua, 35131, Italy}
\newcommand{\PadovaINFN}{Istituto Nazionale di Fisica Nucleare, Sezione di Padova, Padua, 35131, Italy}
\newcommand{\CTU}{Czech Technical University in Prague, Institute of Experimental and Applied Physics, CZ-11000 Prague, Czech Republic}
\newcommand{\Roma}{Universit\`{a} degli Studi di Roma Tre, Rome, 00146, Italy}
\newcommand{\RomaINFN}{Istituto Nazionale di Fisica Nucleare, Sezione di Roma Tre, Rome, 00146, Italy}
\newcommand{\TTU}{Tennessee Tech University, Cookeville, TN 38505, USA}
\newcommand{\NCSU}{Department of Physics, North Carolina State University, Raleigh, NC 27607, USA}	
\newcommand{\SDSMT}{South Dakota Mines, Rapid City, SD, 57701, USA}
\newcommand{\UW}{Center for Experimental Nuclear Physics and Astrophysics, and Department of Physics, University of Washington, Seattle, WA 98195, USA}
\newcommand{\Tuebingen}{University T\"{u}bingen, T\"{u}bingen, 72076, Germany}
\newcommand{\USD}{Department of Physics, University of South Dakota, Vermillion, SD 57069, USA} 
\newcommand{\UZH}{Physik-Institut, University of Z\"{u}rich, Z\"{u}rich, 8057, Switzerland}
\newcommand{\Sudbury}{SNOLAB, Creighton Mine \#9, Sudbury, ON P3Y 1N2, Canada}
\newcommand{\Laurentian}{School of Natural Sciences, Laurentian University, Sudbury, P3E 2C6, Canada} 
\newcommand{\Daresbury}{Science and Technology Facilities Council (STFC) Daresbury Laboratory, Daresbury, Cheshire, WA4 4AD, UK}	%
\newcommand{\Polimi}{Politecnico di Milano, Dipartimento di Elettronica, Informazione e Bioingegneria, Milan, 20133, Italy}
\newcommand{\PolimiINFN}{Istituto Nazionale di Fisica Nucleare, Sezione di Milano, Milan, 20133, Italy}
\newcommand{\LiebnitzPoly}{Leibniz-Institut f\"{u}r Polymerforschung Dresden e.V., Dresden, D-01069, Germany}
\newcommand{\Cagliari}{Istituto Nazionale di Fisica Nucleare (INFN), Sezione di Cagliari, 09042 Italy} %
\newcommand{\UH}{Department of Physics, University of Houston, Houston, TX 77204, USA}
\newcommand{\CIEMAT}{Centro de Investigaciones Energ\'{e}ticas, Medioambientales y Tecnol\'{o}gicas, Madrid, 28040, Spain} % 28040
\newcommand{\Princeton}{Physics Department, Princeton University, Princeton, NJ 08544, USA}
\newcommand{\Frascati}{Istituto Nazionale di Fisica Nucleare, Laboratori Nazionali di Frascati, 00044 Frascati (RM), Italy}
\newcommand{\Naples}{Dipartimento di Fisica dell'Universit\`{a} degli Studi di Napoli ``Federico II'', Naples, 80126, Italy}
\newcommand{\NaplesINFN}{Istituto Nazionale di Fisica Nucleare, Sezione di Napoli, Naples, 80126, Italy}
\newcommand{\Scuola}{Scuola Superiore Meridionale, Naples, 80134, Italy}
\newcommand{\UCSD}{Department of Physics, University of California, San Diego, La Jolla, CA 92093, USA}
\newcommand{\DataSci}{Hal\i c{\i}o\u{g}lu Data Science Institute, University of California, San Diego, La Jolla, CA 92093, USA}
\newcommand{\NTU}{Department of Physics, National Taiwan University, Taipei, 10617, Taiwan}

%author footnotes
%\newcommand{\KCL}{Kings College London, London, United Kingdom}
%\newcommand{\Duluth}{Physics and Astronomy, University of Minnesota, Duluth, MN 55812, USA}
\newcommand{\PNNL}{Pacific Northwest National Laboratory, Richland, WA 99354, USA}
\newcommand{\AFIT}{Air Force Institute of Technology, Dayton, Ohio 45433, USA}
\newcommand{\Henryk}{The Henryk Niewodnicza\'{n}ski  Institute of Nuclear Physics  Polish Academy of Sciences, Krak\'{o}w, Poland}

%%%%%%%%%%%% Need to be in referenced order
\affiliation{\UNC}
\affiliation{\TUNL}
\affiliation{\MPIK}
\affiliation{\UCL}
\affiliation{\SFU}
\affiliation{\UZH}
\affiliation{\USC}
\affiliation{\ORNL}
\affiliation{\UTAustin}
\affiliation{\NRCKI}\thanks{\FootText}
\affiliation{\LNGS}
\affiliation{\Duke}
\affiliation{\Roma}
\affiliation{\RomaINFN}
\affiliation{\INRRAS}\thanks{\FootText}
\affiliation{\GSSI}
\affiliation{\NCSU}	

\affiliation{\UW}
\affiliation{\Polimi}
\affiliation{\PolimiINFN}
\affiliation{\Liverpool}
\affiliation{\PadovaUniv}
\affiliation{\PadovaINFN}
\affiliation{\MPP}
\affiliation{\MILB}
\affiliation{\MILBINFN}
\affiliation{\MILC}
\affiliation{\MILCINFN}
\affiliation{\LBNLENG}
\affiliation{\LBNLNSD}
\affiliation{\SDSMT}
\affiliation{\TUMPhy}

\affiliation{\USD}
\affiliation{\UTK}
\affiliation{\LANL}
\affiliation{\Tuebingen}
\affiliation{\UNM}
\affiliation{\JINR}
\affiliation{\Lancaster}
\affiliation{\IU}
\affiliation{\Frascati}

%the Naples group is too recent to be a L200 author
%\affiliation{\Naples}
%\affiliation{\NaplesINFN}
%\affiliation{\Scuola}

%Regina left March 2025
%\affiliation{\Regina}
\affiliation{\Jag}
\affiliation{\CTU}
\affiliation{\Dresden}
\affiliation{\Geel}
\affiliation{\Bratislava}
%\affiliation{\Sudbury}
%\affiliation{\Laurentian}
\affiliation{\TTU}
\affiliation{\Daresbury}
\affiliation{\MEPhI}\thanks{\FootText}
\affiliation{\UCSD}
\affiliation{\DataSci}
\affiliation{\Queens}
\affiliation{\LAquila}
\affiliation{\Warwick}
\affiliation{\LiebnitzPoly}
\affiliation{\IKZ}

%\affiliation{\NTU}

%%%%%%%%%%%
%LEGEND-1000 only institutions
%%%%%%%%%%%

%\affiliation{\Cagliari}
%\affiliation{\Princeton}
%\affiliation{\CIEMAT}
%\affiliation{\UH}
%\affiliation{\Williams}

%%%%%%%%%%%
% Obsolete institutiokns
%%%%%%%%%%%

%\affiliation{\UCBPH} obsolete
%\affiliation{\IPPFDD} obsolete
%\affiliation{\UCBNE} obsolete

%%%%%%%%%%%
\author{H.~Acharya\,\orcidlink{0009-0000-2219-8511}}\affiliation{\UNC}\affiliation{\TUNL}
\author{N.~Ackermann}\affiliation{\MPIK}
\author{M.~Agostini\,\orcidlink{0000-0003-1151-5301}}\affiliation{\UCL}
\author{A.~Alexander\,\orcidlink{0000-0001-5608-1397}}\affiliation{\UCL}
\author{C.~Andreoiu\,\orcidlink{0000-0002-0196-2792}}\affiliation{\SFU}
\author{G.R.~Araujo\,\orcidlink{0000-0003-4456-0774}}\affiliation{\UZH}
\author{F.T.~Avignone~III\,\orcidlink{0000-0002-5252-9104}}\affiliation{\USC}\affiliation{\ORNL}
\author{M.~Babicz\,\orcidlink{0000-0002-1017-5440}}\affiliation{\UZH}
\author{W.~Bae\,\orcidlink{0000-0002-7646-7577}}\affiliation{\UTAustin}
\author{A.~Bakalyarov}\affiliation{\NRCKI}
% ...

\collaboration{The LEGEND collaboration}
\email[Correspondence: ]{editorial-board@legend-exp.org}
\homepage[Web homepage: ]{https://legend-exp.org}

\newcommand*{\lexpo}{61.0\;\kgyr}
\newcommand*{\lbigold}{0.5\asyerr{0.3}{0.2}\;\cpkty}
\newcommand*{\lbisilv}{1.3\asyerr{0.8}{0.5}\;\cpkty}

\begin{abstract}
  The \legend\ collaboration is searching for neutrinoless double beta (0\nbb) decay by operating
  high-purity germanium detectors enriched in \gesix\ in a low-background liquid argon environment.
  Building on key technological innovations from \gerda\ and the \majoranademo, \legend-200\ has
  performed a first 0\nbb\ decay search based on \lexpo\ of data. Over half of this
  exposure comes from our highest performing detectors, including newly developed \acl*{ic}
  detectors, and is characterized by an estimated background level of \lbigold\ in the 0\nbb\ decay
  signal region. A combined analysis of data from \gerda, the \majoranademo, and \legend-200,
  characterized by a 90\% \acl*{CL} exclusion sensitivity of \smash{$2.8 \times \powtenyr{26}$} on
  the half-life of 0\nbb\ decay, reveals no evidence for a signal and sets a new observed lower
  limit at \smash{$\thalfz > 1.9 \times \powtenyr{26}$} (90\% \acl*{CL}).  Assuming the decay is
  mediated by Majorana neutrinos, this corresponds to an upper limit on the effective Majorana mass
  in the range $\mbb < \text{75--200}$\;meV, depending on the adopted \acl*{nme}.
\end{abstract}

\maketitle

One of the most fundamental open questions in physics is the origin of the matter-antimatter
asymmetry in the universe, observed as an excess of baryons over antibaryons. This imbalance
remains unexplained by the Standard Model of particle physics. A compelling explanation is provided
by the framework of leptogenesis, which proposes that a lepton-antilepton asymmetry generated in
the early universe was partially transformed into the observed baryon asymmetry by electroweak
sphaleron processes~\cite{Fukugita:1986hr}. Neutrinos may play a key role in this scenario,
particularly if they are Majorana fermions~\cite{Majorana:1937vz} --- identical to their
antiparticles --- which inherently violate lepton number conservation~\cite{Schechter:1981bd}.
Neutrinoless double beta (0\nbb) decay, in which two electrons are emitted by an atomic nucleus
without accompanying anti-neutrinos, would directly confirm lepton number violation and the
Majorana nature of neutrinos, marking a revolutionary step in our understanding of the
cosmos~\cite{Agostini:2022zub, Gomez-Cadenas:2023vca, Dolinski:2019nrj}. Additionally, it would
offer insights into a new theory of fermion masses and the absolute neutrino mass scale, bridging
the gap between cosmology and particle physics~\cite{Minkowski:1977sc}.

% Double beta decay with the emission of two anti-neutrinos (2\nbb\ decay) has been observed in several
% nuclei, with half-lives ranging from $10^{19}$ to $10^{24}$\;yr~\cite{Barabash:2020nck}.
A broad experimental program is focused on the search for 0\nbb\ decay in
  \gesix~\cite{Majorana:2022udl,GERDA:2020xhi},
  \nuc{Mo}{100}~\cite{AMoRE:2024loj,Augier:2022znx,Arnold:2015wpy},
  \nuc{Se}{82}~\cite{CUPID:2022puj,Arnold:2018tmo},
  \nuc{Te}{130}~\cite{CUORE:2024ikf,CUORE:2021mvw},
  \nuc{Xe}{136}~\cite{KamLAND-Zen:2024eml,EXO-200:2019rkq},
and other isotopes~\cite{Agostini:2022zub}. The experimental signature of 0\nbb\ decay is a peak in
the spectrum of the summed electron energies at the Q-value (\qbb) of the decay, as the two
electrons carry all the available energy. The \ac{legend} collaboration is pursuing an experimental
program~\cite{LEGEND:2021bnm} with discovery potential at a half-life beyond \powtenyr{28} using
the isotope \gesix\ ($\qbb = 2039.061(7)$\;keV~\cite{Mount:2010zz}). In this letter, we report
results of the first search for \onbb\ decay performed with \legend-200, the initial phase of the
project.

In the first data-taking period, 142.5\;kg of \ac{hpge} detectors
isotopically enriched in \gesix\ to 86--92\% were deployed in \legend-200. \ac{hpge} detectors are
well-suited for double beta decay searches since the energy is deposited locally, as $\sim$1\;MeV
electrons are absorbed within 1--2\;mm of germanium, leading to a detectable, high-resolution signal in a single
detector. The detectors were deployed in the former \gerda~\cite{GERDA:2017ihb} cryostat
and water tank shield at the \ac{lngs} of the Italian \ac{infn}, located at a depth of 3500\;m
water-equivalent.
% The Gran Sasso massif, surrounding the experiment, provides a shielding of 3500\;m
% water-equivalent, reducing the cosmic muon flux by six orders of magnitude~\cite{Borexino:2018pev}.

\begin{table*}
  \newcommand*{\bpoint}{\quad $\triangleright$}
  \newcommand*{\rspan}[1]{\rule[0.5ex]{2.8cm}{0.4pt}\quad#1\quad\rule[0.5ex]{2.8cm}{0.4pt}}
  \caption{%
    Summary of the parameters characterizing the \legend-200 data set presented in this work.
    Energy reconstruction parameters and all 0\nbb\ decay detection efficiencies are reported as
    exposure-weighted averages for each detector type with uncertainties propagated as fully
    correlated. The total mass excludes detectors that could not be operated reliably. The 0\nbb\
    decay detection efficiency of the muon anti-coincidence cut is $>99.9\%$ and is omitted. Data for individual detectors and running periods are provided in~\cite{suppmat}.
  }\label{tab:datapars}

  \begin{ruledtabular}
    \begin{tabular}{lcccc}
      \phantom                                           & \acs*{bege}         & \acs*{ppc}          & \acs*{coax}         & \acs*{ic} \\
      \midrule

      Number of detectors                                & 25                  & 12                  & 6                   & 29 \\
      Total mass                                         & 17.1 kg             & 10.7 kg             & 14.6 kg             & 65.1 kg \\
      Exposure                                           & 10.9 \kgyr          & 4.2 \kgyr           & 7.8 \kgyr           & 38.1 \kgyr \\
      Energy resolution (\acs{fwhm} at \qbb)             & $(2.1 \pm 0.1)$ keV & $(2.5 \pm 0.1)$ keV & $(4.4 \pm 0.1)$ keV & $(2.6 \pm 0.1)$ keV \\
      Energy bias at \qbb\                               & $(0.3 \pm 0.3)$ keV & $(0.3 \pm 0.3)$ keV & $(0.0 \pm 0.3)$ keV & $(0.3 \pm 0.3)$ keV \\
      Signal efficiency                                  & $(49.5 \pm 3.2)$\%  & $(56.1 \pm 3.4)$\%  & $(47.6 \pm 4.5)$\%  & $(59.9 \pm 2.3)$\% \\
      \bpoint\ quality cuts                              & \multicolumn{4}{c}{\rspan{$(97.5 \pm 0.1)$\%}} \\
      \bpoint\ 0\nbb\ decay containment in active volume & $(77.0 \pm 3.2)$\%  & $(83.3 \pm 2.3)$\%  & $(80.4 \pm 4.2)$\%  & $(85.9 \pm 0.8)$\% \\
      \bpoint\ \gesix\ isotopic enrichment               & $(87.4 \pm 0.3)$\%  & $(87.4 \pm 0.5)$\%  & $(86.2 \pm 1.7)$\%  & $(91.3 \pm 0.5)$\% \\
      \bpoint\ pulse shape discrimination                & $(81.1 \pm 3.9)$\%  & $(85.0 \pm 4.5)$\%  & $(76.0 \pm 5.7)$\%  & $(84.3 \pm 3.0)$\% \\
      \bpoint\ liquid argon cut                          & \multicolumn{4}{c}{\rspan{$(93.3 \pm 0.5)$\%}} \\

    \end{tabular}
  \end{ruledtabular}
\end{table*}

The \ac{hpge} detectors are arranged in vertical strings mounted in a circular array 55\;cm in diameter (a rendering is provided in~\cite{suppmat}).
Each detector rests on a scintillating \ac{pen}
plate~\cite{Efremenko:2019xbs, Manzanillas:2022zyh} suspended by rods made of underground
\ac{efcu}~\cite{Hoppe:2014nva}, and is read out by low-radioactivity front-end electronics~\cite{Majorana:2021mtz},
cables, and connectors. The front-ends are affixed to the \ac{pen} plates and connected to charge-sensitive amplifiers operated in
\ac{lar} 30\;cm above the array~\cite{Riboldi:2015dzj}. 
Immersion of the strings into $>$99.999\% pure \ac{lar} in \gerda's 64\;m$^3$ steel cryostat~\cite{Knopfle:2022fso} simultaneously provides cooling, shielding, and background rejection, as the \ac{lar} volume is instrumented to
efficiently detect \acl{vuv} scintillation light produced in the \ac{lar} by background processes.
\Ac{tpb}-coated \ac{wls} fibers, arranged as two
cylindrical curtains, closely surround the \ac{hpge} detector array, and \ac{sipm} modules read out the
guided light at both ends of the fibers~\cite{Costa:2022tsx}. 
A cylindrical \ac{wls} reflector of 138\;cm diameter is
installed concentric to the array~\cite{Araujo:2021buv} to increase light collection.  A dedicated apparatus installed at the
bottom of the cryostat monitors the purity of \ac{lar} over time~\cite{Schwarz:2021xmd}. 
Each string is enclosed in a \ac{wls} \ac{tpb}-coated
nylon cylinder that provides a barrier to \kvz\ ions, produced by \nuc{Ar}{42} decays in
\ac{lar}~\cite{Lubashevskiy:2017lmf}, reducing the \ac{lar} volume from which \kvz\ ions can be
collected on the surface of the \ac{hpge} detectors.
%Four copper funnels mounted at the top of the array are used to guide calibration sources into \ac{tpb}-coated nylon tubes located between the inner fiber curtain and the \ac{hpge} detectors.
The cryostat is deployed in a tank filled with 590\;m$^3$ of purified water, instrumented with 63
\acp{pmt} to tag cosmic-ray induced muon events with $>$99\% efficiency~\cite{Freund:2016fhz}. 
%The \ac{lar} and water surrounding the \ac{hpge} array also constitute a passive shield against trace
%radioactivity in the surrounding materials and laboratory environment. 

% NOTE: moved supplemental material
% \begin{figure}
%   \centering
%   \includegraphics[width=0.8\columnwidth]{l200-setup-paper.pdf}
%   \vspace*{-5mm}

%   \includegraphics[width=0.9\columnwidth]{l200-hpges.pdf}
%   \caption{%
%     Top: rendering of the \legend-200 \ac{hpge} array.  Bottom: artist cutaway view of the four
%     detector types deployed in the first configuration of \legend-200. The weighting potential
%     (color map) and electric field lines at the cutaway plane are overlaid. The bottom bar reports
%     the total deployed mass for each detector type, while the geometry of the \ppluscont\ electrode
%     is highlighted in red.  \acs{bege} and \acs{ppc} types, from the former \gerda\ and the
%     \majoranademo\ experiments, feature small \ppluscont\ electrodes. Similarly for the \legend\
%     \acs{ic} type, which however is more massive.  The \acs{coax} type, inherited from past
%     experiments, is the only one featuring a large \ppluscont\ electrode.
%   }\label{fig:l200-hpges}
% \end{figure}

% hpge detectors
During this first data-taking period, the \ac{hpge} array consisted of 86.7\;kg of
\ic\ detectors~\cite{Cooper:2011ahu, GERDA:2021hud}, along with 22.1\;kg of \ppc\ detectors from the
\majoranademo~\cite{Abgrall:2025tsj}, and 14.7\;kg of \coax\ plus 19\;kg of \bege\ detectors from
\gerda~\cite{GERDA:2019vry}. Typical individual detector masses are 1\;kg for \ppc, 2.5\;kg for
\coax, 0.7\;kg for \bege, and 2--4\;kg for \ic. All are p-type detectors featuring a
lithium-diffused \npluscont\ electrode that wraps around most of the surface and creates a
$\sim$1\;mm inactive layer. A thin boron-implanted \ppluscont\ electrode covers a small fraction of
the surface, except for \coax\ detectors. A thin layer between the electrodes provides electrical
insulation.  The dimensions of the \ppluscont\ electrode and insulating layer typically vary
depending on the manufacturer. An illustration of the general characteristics of the detector
geometry and electric field configuration is provided in~\cite{suppmat}. Detectors produced by
Mirion~\footnote{\url{https://www.mirion.com}} (all \bege\ and 73.4\;kg of \ac{ic}) feature a
smaller insulating surface than those produced by
ORTEC~\footnote{\url{https://www.ortec-online.com}} (all \ppc\ and 13.3\;kg of \ac{ic}). The novel
\ic\ detectors, comprising the largest fraction of the array, feature the most advanced design in
terms of sensitivity to 0\nbb\ decay, combining effective \ac{psd} performance, high energy
resolution and large individual masses to effectively reduce the background level. Ten of the
deployed detectors, totaling about 10\;kg of germanium, could not be operated due to connectivity
issues and were switched off.

% DAQ
The \ac{daq} system synchronously reads out digitized signals from \ac{hpge} detectors and \ac{sipm} modules (16-bit, 62.5\;MHz, 8192 samples)~\cite{puhlhofer_flashcam_2019,diebold_readout_2017}. If an \ac{hpge} signal above a threshold of 25\;keV is
detected, traces from all channels are stored on disk for offline analysis, implemented in the
novel \pygama\ framework~\cite{pygama:v2.0.3, *dspeed:v1.6.1, *daq2lh5:v1.2.1, *lgdo:v1.7.0}.
Signals from the \acp{pmt} (12-bit, 250\;MHz, 220 samples) are acquired independently, with the two
\ac{daq} systems synchronized through a GPS clock.
% For \ac{hpge} channels, only a 22\;\mus\ window around the signal rising edge is kept at full
% resolution, with the rest down-sampled by a factor of eight.
% Additionally, the two least significant bits of the \ac{sipm} signal samples are removed, with
% negligible impact on the accuracy of the reconstructed signals.

% data taking
Twelve \nuc{Th}{228} sources of $\sim$5\;kBq each~\cite{Baudis:2022lcu} are lowered near the
\ac{hpge} detectors for about four hours each week to calibrate their energy scale, resolution, and
\ac{psd} performance. 
%The time between these calibration periods is dedicated to physics data taking.  
A test charge equivalent to a $\sim$1\;MeV signal is injected into the \ac{hpge} front-end
electronics every 20\;s to monitor the stability of the detector response (similar to~\cite{GERDA:2021pcs,Majorana:2021mtz}). Additionally, forced
trigger events (i.e., acquired without a physical signal in the \ac{hpge} detectors) are recorded
at the same rate to study electronic noise and random coincidences. After an initial commissioning
period, a data blinding policy was enforced: events with energy in the \ac{hpge} detectors within
$\pm25$\;keV of \qbb\ were inaccessible until the analysis procedures and parameters were finalized.
% The energy estimator used for blinding is computed online by the \ac{daq} system and calibrated on
% \nuc{Th}{228} data.
% data cleaning
Data from time periods with instrumental instability, due to
temperature fluctuations or hardware interventions, are discarded. 
The data reported here were collected from March 2023 to February 2024, corresponding to a total exposure of 85.5\;\kgyr.

% data processing
% The primary data analysis pipeline is implemented in the novel \pygama\ Python-based
% framework~\cite{pygama:v2.0.2}. A set of \ac{dsp} routines is applied offline to the digitized
% traces to reconstruct various parameters of interest with the \dspeed\
% software~\cite{dspeed:v1.4.0}. A secondary, independent Julia-based framework (\juleana) is
% developed to cross check the results of the main software stack and develop new techniques.

% data quality
A set of quality cuts is applied to identify events incompatible with ordinary energy depositions
in the \ac{hpge} array. These cuts are based on the flatness of the baseline preceding the
\ac{hpge} signal rising edge and the structure of the signal itself, targeting phenomena such as electrical cross-talk or
small electric discharges along the high-voltage line or in the detectors. The estimated acceptance
of physical events at \qbb, reported in \cref{tab:datapars}, varies across detectors and in time.
%Electrical cross-talk between a few pairs of \ac{hpge} and \ac{sipm} channels induces unphysical pulses in the \ac{sipm} trace; a dedicated cut based on the peculiar shape of such signals excludes these traces from the analysis. 

% energy scale and resolution
%Precise control of energy scale stability and achieving high energy resolution are fundamental to maximizing sensitivity to 0\nbb\ decay. 
The energy of \ac{hpge} events is reconstructed using a
truncated cusp filter~\cite{Gatti:1986cw}, with parameters optimized for each detector and
data-taking run based on weekly \nuc{Th}{228} calibration data.
% We compared the energy resolution obtained using different methods and found that the cusp filter
% provides the best performance compared to standard trapezoidal and ZAC~\cite{GERDA:2015rik}
% filters.
A charge-trapping correction is also applied to the energy values~\cite{Majorana:2022vai}. The
observed energy resolution (see \cref{tab:datapars}) is improved by 20--30\% compared to that
achieved in \gerdatwo. %, thanks to the front-end electronics developed by the \majorana\ collaboration.
The energy calibration procedure follows prior work~\cite{GERDA:2021pcs, Majorana:2023kdx}. The
stability of the energy scale and resolution is monitored over time using calibration data. Data
from detectors and periods with unstable energy scale (e.g.~when instabilities at \qbb\ are
comparable to the energy resolution), amounting to 11.3\;\kgyr, are only used to determine event
multiplicity, i.e., number of detectors in which a signal was recorded. The bias of the energy
estimator at \qbb\ due to ADC non-linearity is estimated following~\cite{GERDA:2021pcs} as the
difference between 2039.06\;keV and the estimated energy (see \cref{tab:datapars}) and is accounted
for in the 0\nbb\ decay analysis.

% multiplicity cut and muon veto
Events with energy depositions in multiple \ac{hpge} detectors are rejected. The evaluation of
event multiplicity accounts for electrical cross-talk between channels (following~\cite{WesterPhD2020}), which is estimated from
calibration data to be below 1\% for most cases, with only a few channel pairs exceeding this
threshold.  Data from \acp{pmt} in the water tank are analyzed offline to identify muon signals.
\Ac{hpge} events accompanied by a muon signal in a window of 3.5\;\mus\ around an \ac{hpge} trigger
are also discarded with a negligible dead time.

\begin{figure*}
  \includegraphics[width=\textwidth]{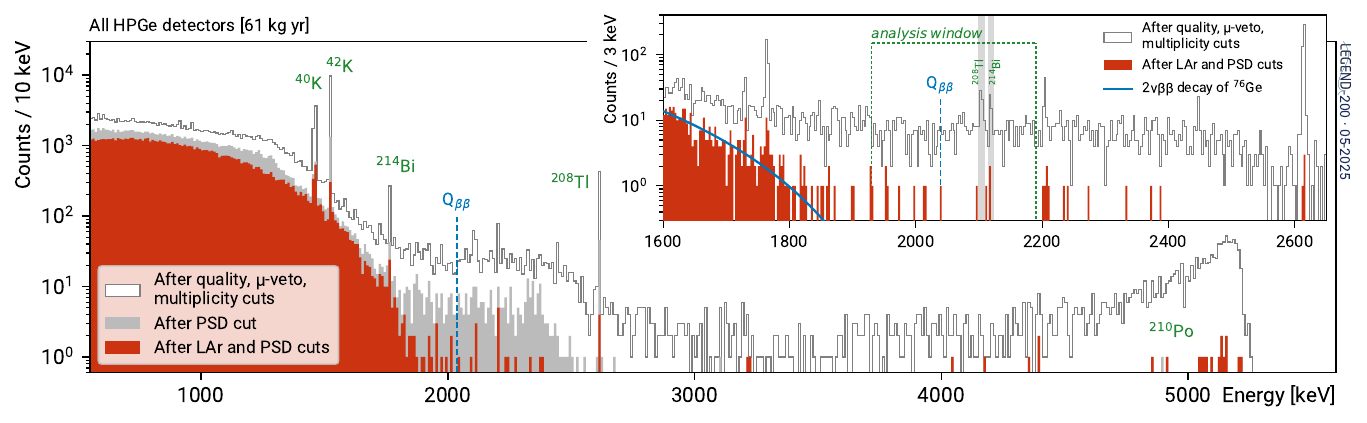}
  \caption{%
    The energy spectrum of the first \legend-200 data set, corresponding to \lexpo\ of germanium
    exposure, above the \Arl\ $\qb = 565$\;keV. The white histogram shows events passing quality
    and muon anti-coincidence cuts and with energy deposited in one single \ac{hpge}
    detector. The main radioactive background contributors are indicated in green. The grey and red
    histograms show the subsets of events passing the \acs{psd} cut and additionally the \acs{lar} anti-coincidence cut, respectively. The inset shows a close-up around the region of interest for 0\nbb\ decay with
    finer binning. The expected contribution from the 2\nbb\ decay of \gesix\ (\smash{$\thalft =
    2.022 \times \powtenyr{21}$}~\cite{GERDA:2023wbr}) corresponds to the solid blue line. The
    events used to set a constraint on the 0\nbb\ decay rate are contained by the analysis window,
    marked in green. Only two significant \g\ peaks (shaded areas) are expected within this window,
    and the corresponding 10\;keV wide energy regions are excluded from the statistical analysis.
  }\label{fig:l200-spectrum}
\end{figure*}

% background model
The physics data collected by \legend-200 surviving quality, \ac{hpge} multiplicity, and muon cuts
are shown in \cref{fig:l200-spectrum}.  The types of background events observed are the same as
those in \gerda~\cite{GERDA:2019cav}.  We characterize the dominant background contributions and
validate the interpretation of spectral features by comparing data and \ac{MC} simulations, following the same techniques used in \gerda~\cite{GERDA:2019cav} and the \majoranademo~\cite{Reine:2023ztz,Haufe:2023tux}.  At
energies below \qbb, the spectrum is dominated by the continuous distribution of 2\nbb\ decay
events. The \g\ lines and continuum below $\sim$3\;MeV originate from \g\ rays emitted by the \kvn,
\nuc{U}{238}, and \nuc{Th}{232} decay chains in the structural materials, and also the decay of
\nuc{K}{42} in \ac{lar}. These \g\ rays can be fully absorbed within a single \ac{hpge} detector
or may deposit some energy in germanium and the surrounding optically active materials, primarily
\ac{lar} and \ac{pen}. In the first case, multiple Compton scatters can occur within an \ac{hpge}
detector (\ac{mse}), which differs from the \ac{sse} topology of \b\b\ decay. Even if
only a single-site Compton scatter takes place in an \ac{hpge} detector, coincident scintillation
light due to \g\ or accompanying \b\ radiation is likely to be detected.  At higher energy the
spectrum is dominated by energy-degraded \a\ particles from the \nuc{U}{238} decay chain, primarily
from \nuc{Po}{210} decays on the \ppluscont\ electrode or insulating surfaces. The vast majority of
these \a\ events originate from the \coax\ detectors, featuring the largest \ppluscont\ surfaces.
Additionally, \b\ particles from \kvz\ decay ($\qb = 3.53$\;MeV) on the detector surface can
penetrate into the active volume.

% LAr veto
To evaluate the presence of scintillation light coincident with the \ac{hpge} signal,
\ac{sipm} traces are analyzed to reconstruct the time and amplitude of each pulse.
Low-amplitude pulses from transient noise in the \acp{sipm} are discarded, with a
rejection threshold varying between channels, typically around 0.5 \acp{pe}. The coincidence window
is defined as $[-1, 5]$\;\mus\ relative to the onset of the \ac{hpge} rising edge. Events are
excluded from the analysis if the sum of the amplitudes across all \ac{sipm} channels exceeds four
\acp{pe}, or if the multiplicity (i.e., the number of channels with a signal above threshold)
exceeds four, considering only pulses within the coincidence window.  This cut results in a
survival fraction of signal events of $(93.3 \pm 0.5)$\%, as determined using forced trigger events
and \kvn\ \ac{fep} events, where no coincident light is expected.
% The higher threshold compared to \gerda~\cite{GERDA:2020xhi} arises from the improved light
% collection efficiency of the \legend-200 \ac{lar} instrumentation, which covers a larger volume
% around the array and leads to a higher rate of random coincidences from \nuc{Ar}{39} decays in
% \ac{lar}.

% PSD cut
The shape of the rising edge of \ac{hpge} signals is analyzed to further identify background
events.  We use two \ac{psd} techniques to identify signal-like event topologies in detectors
featuring a small \ppluscont\ electrode: first, \aoe\ measures the maximum current amplitude ($A$) over
the charge amplitude ($E$)~\cite{GERDA:2022ixh, Majorana:2019ftu}. Second, \ac{lq} measures the
area above the last 20\% of the charge signal normalized by the charge
amplitude~\cite{Majorana:2022udl}.  For signal-like events, these parameters are normally
distributed, with mean and standard deviation measured using the 1593\;keV \ac{dep} induced by the
2614\;keV \nuc{Tl}{208} \g\ ray from the calibration sources, a \ac{sse}-enriched event sample. Both
\ac{psd} estimators are corrected for correlations with drift time~\cite{Comellato:2020ljj} and
energy.

\aoe\ is used to reject \acp{mse}, which exhibit lower \aoe\ values than \acp{sse}. The cut
threshold is tuned to achieve 90\% survival fraction of \nuc{Tl}{208} \ac{dep} events. \a\ and
\b\ particles incident on surfaces can also be rejected by \ac{psd}.  Events near the \ppluscont\
electrode result in high values of \aoe\ and \ac{lq}. Events near the \npluscont\ and passivated
surfaces produce low values of \aoe\ and high values of \ac{lq}. For Mirion \ic\ and \bege\ detectors
with a narrow passivated groove between the \ppluscont\ and \npluscont\ electrode, a value of \aoe\
three standard deviations above the mean is used to reject these events. For Ortec detectors with a wide
passivated surface (\acp{ppc} and one \ac{ic}), a value of \ac{lq} three standard deviations above
the mean is used. Five \ic\ detectors with abnormally wide \ppluscont\ pads (indicated as \icfive)
had a significant population of \ac{mse} that passed the \aoe\ cut.  For these detectors \ac{lq}
values three standard deviations above the mean are also used to reject \acp{mse}. The \coax\
detectors exhibit different pulse-shape characteristics than detectors featuring a small
\ppluscont\ electrode, severely reducing the effectiveness of \aoe\ and \ac{lq} cuts.  For these detectors, following
the procedure in~\cite{GERDA:2022ixh}, an artificial neural network is employed to reject 
\acp{mse}, in combination with a rise time cut to eliminate fast events originating from the
\ppluscont\ electrode.

The 0\nbb\ decay PSD survival fraction is corrected for several systematic effects. We correct for
dependence on energy, measured using a \nuc{Co}{56} source with multiple \acp{dep} ranging from
1013 to 2429\;keV~\cite{Majorana:2022udl}. We also correct for the differences in spatial
distribution between \ac{dep} events (concentrated near surfaces) and \b\b\ events (uniformly
distributed in the bulk) by measuring the detection efficiency for 2\nbb\ events in an energy range
of 1.0--1.3\;MeV.  These corrections reduce the 0\nbb\ decay detection efficiency (reported in
\cref{tab:datapars}) by a few percent. Finally, we add uncertainties related to weekly shifts in
the \ac{dep} efficiency and energy dependence.  Data from periods characterized by significantly
unstable \ac{psd} performance, amounting to 13.2\;\kgyr, are only used to determine event
multiplicity.

After data selection, the total exposure used for analysis amounts to \lexpo. The
energy spectrum, after applying the \ac{lar} anti-coincidence and \ac{psd} cuts, is shown in red
in \cref{fig:l200-spectrum}. A remarkably low background level is observed across the entire energy
range, above the 2\nbb\ decay events. At \qbb, the \ac{lar} cut provides the primary rejection
of Compton scatters from the 2615\;keV \g\ ray of \nuc{Tl}{208}, which dominate the spectrum prior
to cuts. The \ac{psd} cut independently removes about 60\% of these events, as measured in
calibration data.  The next largest contributors, \b\ decays of \nuc{K}{42} on the \ac{hpge}
surface and \g\ rays from \nuc{Bi}{214}, are also strongly suppressed by the cuts. The enhanced
rejection achieved through the combination of \ac{lar} and \ac{psd} cuts reflects the
complementarity of these techniques, which target different interaction topologies 
%(see \cref{fig:l200-result}, bottom panel). 
(see \cref{fig:l200-spectrum}). 
The contribution of cosmogenic backgrounds is estimated to
be less than $2 \times \powcpkky{-5}$~\cite{Wiesinger:2018qxt}.

\begin{figure}
  \includegraphics[width=\columnwidth]{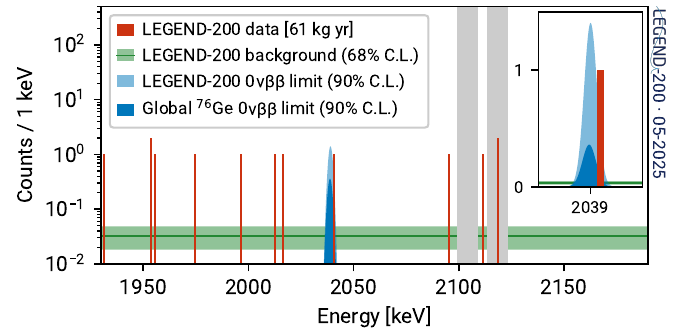}
  \caption{%
    %Top: the 
    The energy spectrum of the first \legend-200 data set (\lexpo) after all analysis cuts in
    the $[1930, 2190]$\;keV window. Events in the gray regions, corresponding to $\pm5$\;keV around
    known \g\ lines, are excluded from the analysis. Confidence intervals from the frequentist
    analysis are visualized for the exposure-weighted combined background index (68\% \acs*{CL}, in green) and for the signal
    strength (90\% \acs*{CL}, in blue). The 0\nbb\ decay rate upper limit derived by including data
    from other \gesix\ experiments is shown in dark blue. 
    %Bottom: distribution of \aoe\ classifier
    %and number of coincident \aclp{pe} in \ac{lar} for events from \ic\ Mirion detectors in the
    %same energy window as the top panel. The \aoe\ classifier is calibrated such that its
    %distribution for single-site events is approximately normal, centered at zero with a standard
    %deviation of 1.
}\label{fig:l200-result}
\end{figure}

% statistics
Statistical inference on the 0\nbb\ decay rate is performed using both frequentist and Bayesian
frameworks. Following~\cite{GERDA:2020xhi}, the analysis window is defined as the energy range from 1930\;keV to 2190\;keV,
excluding $\pm 5$\;keV regions around the 2104\;keV (\nuc{Tl}{208}) and 2119\;keV (\nuc{Bi}{214})
\g\ ray \acp{fep}. There are eleven events that survive all cuts within this window (see
\cref{fig:l200-result}). The low event count and the expected background flatness in the region of
interest allow the use of an analytical model comprising a uniform component and a normal
distribution at the expected 0\nbb\ decay signal location. The likelihood function is defined as
the product of extended unbinned likelihood terms for each \ac{hpge} detector and for each time
period (``partitions'', see detailed description in~\cite{Agostini:2017iyd}) with stable energy
reconstruction and 0\nbb\ decay detection efficiency parameters.

The location, scale, and normalization parameters of the normal distribution modeling 0\nbb\ decay
in each likelihood term are constrained to their expected values using normal distributions as pull
terms (or priors in the Bayesian approach). The uncertainty on the 0\nbb\ decay detection
efficiency is conservatively assumed to be maximally correlated between partitions, simplifying the
model by reducing the number of free parameters. If a statistical test does not indicate
significant variation of background rate within a data set, a single background index parameter is
used to model the background across all its partitions.  Following this approach, data from \coax\
and \icfive\ detectors (the \silver\ data set, 12.7\;\kgyr) are assigned a separate background
index, as their background level is expected to be higher than that of the rest of the data (the
\golden\ data set, 48.3\;\kgyr). All information required to implement the likelihood function
used here is provided in~\cite{LEGEND:zenodosupp}.

The frequentist analysis~\cite{freqfit:v0.2} uses a two-sided profile likelihood ratio test
statistic with parameters constrained to physical values~\cite{Cowan:2010js}; its distribution is
computed via \ac{MC} methods. No evidence of a signal is found, resulting in a lower limit on
the half-life of 0\nbb\ decay of $0.5 \times \powtenyr{26}$ at 90\% \ac{CL}. The background index
at 68\% \ac{CL} is \lbigold\ for the \golden\ data set and \lbisilv\ for the \silver\ data set; both are consistent within uncertainty with the \gerda\ background index~\cite{GERDA:2020xhi}.
A visualization of these confidence intervals is shown in \cref{fig:l200-result}. The Bayesian
analysis is performed using the \texttt{BAT.jl} software library~\cite{ZeroNuFit.jl:v2.3.1,
Schulz:2021BAT, *BAT.jl:v3.3.4}, adopting uniform priors for the signal and background rates. The
resulting 90\% \ac{CI} on the signal strength is identical to that obtained in the frequentist
analysis.

The expected statistical distribution of 90\% \ac{CL}~intervals was estimated via toy \ac{MC}
experiments under the assumption of no signal. The observed limit lies within the central 95\%
interval of the expected distribution and is less stringent than its median of $1.0 \times
\powtenyr{26}$ (see~\cite{suppmat}). This is partly due to an event at an energy 1.3 standard
deviations from the expected signal location within its partition. The event was recorded in a
\ppc\ detector and exhibits a pulse shape with a high value of A/E, compatible with an interaction near the \ppluscont\
electrode.

The full \majoranademo~\cite{Majorana:2022udl} and \gerda~\cite{GERDA:2020xhi} data
sets (amounting to 64.5 \kgyr\ and 127.2 \kgyr\ of exposure, respectively) are incorporated into
the analysis as additional extended terms in the likelihood to extract a combined 0\nbb\ decay
constraint. These terms are unchanged from the original analyses.  With a p-value of 0.29
for the background-only hypothesis in the combined data set, no evidence of a signal is found.
Therefore, we extract a combined lower limit on the half-life of 0\nbb\ decay of $1.9 \times
\powtenyr{26}$ (90\% \ac{CL}). The corresponding 90\% \ac{CL} interval for the signal strength in
the \legend-200 data set is visualized in \cref{fig:l200-result}. The observed limit lies within
the central 68\% interval of the expected limit distribution, characterized by a median of $2.8
\times \powtenyr{26}$ (see~\cite{suppmat}), the best achieved among 0\nbb\ decay searches to date.
The Bayesian analysis yields an identical 90\% \ac{CI} lower limit under uniform signal and
background priors. Adopting a uniform prior on the Majorana neutrino
mass yields a stronger limit of $2.8 \times \powtenyr{26}$.
%, stronger than with a uniform prior as expected in the absence of a signal.

The overall effect of uncertainties in the 0\nbb\ decay signal model on the limit is at the percent
level. Systematic uncertainties in the fit model have a marginal impact; for example, assuming a
linear background shifts the limit by a few percent.

We calculate constraints on the effective Majorana mass \mbb\ using a range of \acp{nme}  calculations~\cite{% list copied from https://arxiv.org/abs/2212.11099
  Menendez:2017fdf,%
  %Horoi:2015tkc,% no Ge #'s
  Coraggio:2020hwx,%
  %Coraggio:2022vgy,% no Ge #'s
  Mustonen:2013zu,%
  Hyvarinen:2015bda,%
  Simkovic:2018hiq,%
  Fang:2018tui,%
  %Terasaki:2020ndc,% no Ge #'s
  Rodriguez:2010mn,%
  LopezVaquero:2013yji,%
  Song:2017ktj,%
  Barea:2015kwa,%
  Deppisch:2020ztt,%
  Jiao:2017opc%
}, i.e.~2.35--6.34, yielding a range of upper limits of $\mbb < \text{75--200}$\;meV in the
frequentist framework. In addition, we provide a second estimate based on a recent
\textit{ab-initio} calculation
that includes "quenching" physics~\cite{Gysbers:2019uyb}, 
the leading short-range neutrino-exchange mechanism~\cite{Cirigliano:2018hja,Cirigliano:2020dmx}, 
and, for the first time, a comprehensive Bayesian treatment
of theoretical uncertainties~\cite{Belley:2023lec}. Using this approach, we derive an upper limit
of $\mbb < 320$\;meV in the Bayesian framework, with its strength significantly limited by the
large uncertainty in the \ac{nme} value. The posterior distributions are available
in~\cite{suppmat}.

% >>> comparison with other experiments
% - KamLAND-Zen: limit 2.3E26, sensitivity 1.5, mbb < 36--156 meV
% - KamLAND-Zen Nu24 (but pre-print): limit 3.8E26, sensitivity 2.6, mbb < 28--122 meV

% Among the experiments searching for neutrinoless double beta decay, \kamlandzen\ has set some of
% the most stringent constraints, with a reported half-life limit of $2.3 \times 10^{26}$~yr and a
% sensitivity of $1.5 \times 10^{26}$~yr~\cite{KamLAND-Zen:2022tow}. The results presented in this
% letter surpass these limits, although a new KamLAND-Zen analysis with more stringent constraints is
% currently under peer review~\cite{KamLAND-Zen:2024eml}.

We have presented first results from the initial data-taking phase of \legend-200. With the
deployment of additional large mass \ac{ic} detectors and background reduction through refined
surface treatment of nearby components, data acquisition will resume with an improved detector array.
This paper marks the beginning of the phased \legend\ program, which ultimately aims to operate up
to one ton of \ac{hpge} detectors enriched in \gesix\ in the future \legend-1000
infrastructure~\cite{LEGEND:2021bnm}.  The program is designed to achieve discovery sensitivities
for 0\nbb\ decay half-lives beyond $10^{28}$ years that explore the inverted and a significant
fraction of the normal neutrino mass ordering regime, as predicted by neutrino oscillation
experiments~\cite{Strumia:2006db, Agostini:2022zub}.

\begin{acknowledgments}
  This material is based upon work supported by the U.S.~Department of Energy, Office of Science,
  Office of Nuclear Physics under Federal Prime Agreements  DE-AC02-05CH11231, DE-AC05-00OR22725,
  LANLEM78, and under award numbers DE-SC0017594, DE-FG02-97ER41020, DE-FG02-97ER41033,
  DE-FG02-97ER41041, DE-FG02-97ER41042, DE-SC0017594, DOE DE-SC0022339, DE-SC0012612, DE-SC0018060,
  and DE-SC0014445.  We acknowledge support from the Nuclear Precision Measurements program of the
  Division of Physics of the National Science Foundation through grant numbers NSF PHY-1812374, NSF
  PHY-1812356, NSF-PHY-2111140 NSF PHY-1812409, NSF PHY-2209530, NSF PHY-2312278, and from the
  Office of International Science and Engineering of the National Science Foundation through grant
  number NSF OISE 1743790.  We gratefully acknowledge the support of the U.S.~Department of Energy
  through the LANL, ORNL and LBNL Laboratory Directed Research and Development (LDRD) Programs for
  this work.  This research is funded in part by the Deutsche Forschungsgemeinschaft (DFG, German
  Research Foundation) -- Excellence Cluster ORIGINS EXC 2094-39078331; SFB1258-283604770.  We
  acknowledge the support of the German Federal Ministry for Education and Research (BMBF) through
  grant number 05A2023.  and the Max Planck Society (MPG).  This work is supported in part by the
  European Research Council (ERC) under the European Union's Horizon 2020 research and innovation
  programme (Grant agreement No. 786430 -- GemX).  We gratefully acknowledge the financial support
  of the Italian Istituto Nazionale di Fisica (INFN), the Polish National Science Centre (NCN,
  grant number UMO-2020/37/B/ST2/03905), the Polish Ministry of Science and Higher Education
  (MNiSW, grant number DIR/WK/2018/08 and 2022/WK/10), the Czech Republic Ministry of Education,
  Youth and Sports LM2023063, the Slovak Research and Development Agency, grant APVV-21-0377, and
  the Swiss National Science Foundation (SNF), SNF FLARE 20FL20\_216572, and FLARE 20FL20\_232670,
  and SNF 200020\_219290.  This project has received funding /support from the European Union's
  Horizon 2020 research and innovation programme under the Marie Sk\l{}odowska-Curie grant
  agreement No 860881-HIDDeN.  This work has been supported by the Science and Technology
  Facilities Council (STFC), part of U.K.~Research and Innovation  (grant numbers ST/W00058X/1 and
  ST/T004169/1).  We acknowledge the support of the Natural Sciences and Engineering Research
  Council of Canada, funding reference number SAPIN-2017-00023.  This research used resources
  provided by National Energy Research Scientific Computing Center (NERSC), a U.S. Department of
  Energy Office of Science User Facility at LBNL, and the Oak Ridge Leadership Computing Facility
  at Oak Ridge National Laboratory.  We thank the directors and the staff of the Laboratori
  Nazionali del Gran Sasso and our colleagues at the Sanford Underground Research Facility for
  their continuous strong support of the LEGEND experiment.
  We would like to thank the authors of \cite{Belley:2023lec} for providing the posterior
  distribution of the \ac{nme} for the 0\nbb\ decay of \gesix\ considered in this work.
\end{acknowledgments}

\bibliography{references.bib}

\end{document}